\documentstyle[12pt,amssymb,amsfonts,epsfig]{article}

\def\d{\partial}
\def\l{\left(}
\def\r{\right)}\def\e{\mbox{e}}

\begin{document}
\begin{center}
  {\Large\bf Is the electric charge conserved in brane world?} \\
  \medskip S.~L.~Dubovsky$^a$, V.~A.~Rubakov$^a$, P.~G.~Tinyakov$^{a,b}$\\
    \medskip
  {\small
     $^a$Institute for Nuclear Research of
         the Russian Academy of Sciences,\\  60th October Anniversary
  Prospect, 7a, 117312 Moscow, Russia\\
$^b$Institute of Theoretical Physics,
University of Lausanne, CH-1015 Lausanne, Switzerland\\
  }

\begin{abstract}
We discuss whether electric charge conservation may not hold in
four-dimensional world in models with infinite extra dimensions, i.e.,
whether escape of charged particles from our brane is consistent with
effectively four-dimensional electrodynamics on the brane. We
introduce a setup with photon localized on the brane and show that
charge leakage into extra dimension is allowed within this setup. The
electric field induced on the brane by escaping charge does not obey
four-dimensional Maxwell's equations; this field gradually disappears
in a causal way. We also speculate on the possibility of the escape of
colored particles and formation of colorless free quark states on the
brane.
\end{abstract}
\end{center}

The discussion of whether electric charge may not be exactly conserved
has long history \cite{Okun1,Okun,Ignat,all}. In four-dimensional theories,
even tiny non-conservation of electric charge leads to contradictions
to low-energy tests of quantum electrodynamics unless exotic
millicharged particles are introduced \cite{Ignat} (see, however,
Ref.~\cite{Ignat1}). A new perspective
emerges in theories describing our world as a brane embedded in higher
dimensional space with infinite extra dimensions
\cite{RShap,Akama,DvaliShifman,RanSun}.
It is conceivable that in these theories, particles initially
residing on our brane may eventually leave the brane and disappear
into extra dimensions. In fact, this leakage of particles has been
found to be generic at least in a class of field theoretic models of
localization of matter on a brane: once matter fields get small
but non-vanishing masses, the localization becomes incomplete and
particles tunnel from the brane into extra dimensions \cite{DRT}.

If particles that leave our brane are electrically charged, their
disappearance into extra dimensions would result in the
non-conservation of electric charge, as seen by a four-dimensional
observer. Of course, in this scenario electric charge is conserved in
the full multi-dimensional space; the charge non-conservation in
four-dimensional world is merely a consequence of incapability of our
devices to detect the charges outside the brane. Still, this scenario
has distinctive experimental signatures such as literally disappearing
electrons. Hence, an observation of processes like $\e^-\to nothing$
would be a strong evidence for the existence of infinite extra
dimensions.  In a sence, this probe of extra dimensions is
complementary to searches expoloiting the possibility of low
fundamental gravity scale: the latter possibility exists
irrespectively of whether extra dimensions are compact \cite{ADD,AADD,RanSun2b}
or infinite \cite{LykRan} whereas the disappearance of matter emerges
in theories with infinite extra dimensions but does not require the
low fundamental scale of gravity.  The rates of proccesses like
$\e^-\to nothing$ are naturally small \cite{DRT} but presently cannot
be reliably predicted as they depend on the localization mechanism and
unknown parameters of extra-dimensional physics.

One may worry that non-conservation of electric charge in
our world may be in contradiction with the fact that electrodynamics
on the brane is effectively four-dimensional. Indeed, the standard
lore is that the charge conservation in our world is guaranteed by the
four-dimensional Gauss' law (and causality, i.e., absence of
action-at-a-distance). The purpose of this paper is to  prove by
example that there is no contradiction at all: in a model we consider,
escape of charges into
extra dimensions is perfectly consistent with localization of
electromagnetism on the brane. The resulting picture is that physics
becomes intrinsically multi-dimensional once the charged particles
leave the brane; the four-dimensional Gauss' law is no longer valid;
the electromagnetic field induced on the brane by escaping  charges
does not obey the four-dimensional Maxwell's equations; this induced
field gradually decreases in a causal manner and finally disappears.

The issue we discuss in this paper has a close gravitational analogy:
matter leaking into extra dimensions carries away energy, so one may
wonder whether this is consistent with the ``Gauss' law'' of
four-dimensional general relativity describing gravity on our
brane. The gravitational field of escaping particles has been studied
in Ref. \cite{GRS2} with the results very similar to those outlined in
the previous paragraph. We will closely follow Ref. \cite{GRS2} both
in spirit and in some of the technicalities.

To discuss the consistency of charge non-conservation on the brane
with four-dimensional behavior of electromagnetism, we
have in the first place to construct a model with a localized photon. 
A simple possibility is the localization of a photon  due to
gravity produced by the brane itself, which is very much the same
mechanism as one leading to the localization of a graviton
\cite{RanSun} and a scalar \cite{Giga}. Consider space-time with
$(4+n+1)$ dimensions, $n$ of which are compact and one is infinite and
``warped''. For simplicity we consider compactification on
$n$-dimensional torus. Let our world be a $(3+n)$-brane with $n$
compact dimensions of very small size (smaller than TeV$^{-1}$). The
brane has a certain tension, and there is a negative bulk
cosmological constant. With this setup and appropriate fine-tuning
between the bulk cosmological constant and brane tension, there exists
a solution to $(4+n+1)$-dimensional Einstein equations with metric
similar to the Randall--Sundrum \cite{RanSun} one,
\begin{equation}
\label{*}
ds^2=a^2(z)\left[\eta_{\mu\nu}dx^{\mu}dx^{\nu}-\sum_{i=1}^nR_i^2d\theta_i^2
\right]-dz^2\;,
\end{equation}
where $\eta_{\mu\nu}$ is the four-dimensional Minkowski tensor,
$\theta_i\in [0,2\pi]$ are compact coordinates and $R_i$ are (small)
sizes of compact dimensions. Here
\[
a(z)=\e^{-k|z|}
\]
and $k$ is determined by the bulk cosmological constant.

Let us now introduce the $U(1)$ gauge field $A_M$ propagating in the 
background (\ref{*}). With appropriate rescaling, its free action is
\[
S_{gauge}=-{1\over 4}\int dz\cdot\prod{d\theta_i\over 2\pi R_i}\cdot
d^4x \sqrt{g}g^{MP}g^{NQ}F_{MN}F_{PQ}\;.
\]
As we consider small $R_i$ and low energies, we truncate this action
to the zero Kaluza--Klein modes of compact $n$ dimensions, i.e., take
$A_M$ independent of $\theta_i$. To see that there exists a photon
localized on the brane, consider the action for four-vector components
$A_{\mu}(x,z)$ of the gauge potential,
\[
S=-{1\over 4}\int dza^n\eta^{\mu\nu}\eta^{\lambda\rho}F_{\mu\lambda}F_{\nu\rho}+\dots
\]
We see that the measure determining the normalization factor is
\begin{equation}
\label{3+}
\int dz a^n\equiv \int_{-\infty}^{\infty}dz\e^{-kn|z|}\;.
\end{equation}
Hence, the $z$-independent mode, $A_{\mu}(x)$, is normalizable; this
is the wave function of the localized photon, up to a normalization
factor proportional to $\sqrt{kn}$. The reason of why we had to
generalize the five-dimensional Randall--Sundrum setup is now obvious: at
$n=0$ there is no localized photon \cite{Giga}, but at $n\geq 1$ gauge
fields localize on the brane, and
electrodynamics on the brane becomes four-dimensional at large
distances.

The main purpose of this paper is to calculate the gauge field induced
on the brane by charges escaping into the non-compact dimension. We
again take the charged fields independent of $\theta_i$. The motion of
charges along the $z$-direction is then treated in the classical
approximation. Let us consider for definiteness a single particle that
has been at rest on the brane at ${\bf x}=0$ and $t<0$ and
then (at time $t=0$) escapes the brane towards positive
$z$ with zero initial velocity. (The same calculation applies to the
process of escape of a particle in an orbifold setup with fixed point $z=0$.)

The particle moves perpendicular to the
brane and gets accelerated towards large $z$ by the gravitational
field \cite{Volovich,GRS2}. The world line of this particle is given
by
\begin{equation}
\label{traj}
{\bf x}_c=0;\;z_c(t)={1\over 2k}\mbox{ln}(1+k^2t^2)\;.
\end{equation}
For a particle of charge $e$, the non-vanishing components of the
corresponding electromagnetic current are
\begin{eqnarray}
\label{current}
\sqrt{g}j^{0}=e\delta^{(3)}({\bf x})\delta(z-z_c(t))\\
\sqrt{g}j^{z}=e\delta^{(3)}({\bf x})\delta(z-z_c(t)){dz_c\over dt}\;.
\end{eqnarray}
Hereafter $g=|\det{g}|$.
This current induces the electromagnetic field everywhere in space-time
according to Maxwell's equations
\begin{equation}
\label{max}
\d_M\left(\sqrt{g}g^{MN}g^{PQ}F_{NQ}\right)=-\sqrt{g}j^{P}\;.
\end{equation}
Since $j^{\theta_i}=0$, it is consistent to set $A_{\theta_i}=0$, and
also consider $A_M$ independent of $\theta_i$. Furthermore, we can
choose the gauge
\[
A_z=0\;.
\]
Then Maxwell's equations (\ref{max}) reduce to two equations. On the
right to the brane these are
\begin{eqnarray}
\label{nu}
\e^{-nkz}\d_{\mu}F^{\mu\nu}-\d_z\left(\e^{-(n+2)kz}\d_zA^{\nu}\right)=
-\delta^{\nu0}\sqrt{g}j^0\\
\label{z}
\e^{-(n+2)kz}\d_z\d_{\mu}A^{\mu}=-\sqrt{g}j^z
\;,
\end{eqnarray}
where the four-dimensional indices are raised and lowered by the
Minkowski metric.

As long as we are interested in the electromagnetic field
{\it on the brane}, only Eq.~(\ref{nu}) is important. This
equation (again on the right to the brane for definiteness)
can be written in the following way
\begin{eqnarray}
\label{6*}
\e^{-nkz}\left[\d_{\mu}^2-\e^{-2kz}\d_z^2+(n+2)k\e^{-2kz}\d_z\right]A^{\nu}=
-\delta^{\nu0}\sqrt{g}j^0+\e^{-nkz}\d^{\nu}\l\d_{\mu}A^{\mu}\r\;.
\end{eqnarray}
The last term in this equation can be explicitly found by making use
of Eq. (\ref{z}), but it produces pure gauge contribution 
{\it on the brane}. Indeed, the solution to Eq.~(\ref{6*}) is
expressed in terms of the retarded Green's function, $G_R(x-x',z,z')$,
of the operator entering the left hand side,
\begin{eqnarray}
A^{\nu}(x,z)=-\int d^4x'dz' G_R(x-x',z,z')\delta^{\nu
0}\sqrt{g}j^0(x',z')\nonumber\\ +\d_{\nu}\int d^4x'dz'
G_R(x-x',z,z')\d_{\mu}A^{\mu}(x',z')\e^{-nkz'}\;.\nonumber
\end{eqnarray}
The second term here is pure gauge on the brane; it does not affect
charges residing on the brane and hence can be omitted. The only
relevant component of $A^{\mu}$ on the brane is then
\[
A^0(x)\equiv A^0(x,z=0)=-\int
d^4x'dz'G_R(x-x',0,z')\sqrt{g}j^0(x',z') \;.
\]
Upon introducing a variable
\[
\xi={1\over k}\e^{kz}
\]
this expression is written in the following explicit form
\begin{equation}
\label{4*}
A^0(r,t)=-\int dt'\int_{1/k}^\infty d\xi'
G_R\l t-t',r,\xi ={1\over k},\xi'\r
e\delta(\xi'-\xi_c(t'))\;.
\end{equation}
We will be interested in length scales much larger than $k^{-1}$, so
we will set
\[
\xi_c(t)=t
\]
in the rest of our analysis.

 To proceed further, we need an explicit
form of the retarded Green's function,
\begin{equation}
\label{summode}
G_R(x-x',z,z')=\left[kn D_0(x-x')+\int_0^{\infty}dm
A_m(z)A_m(z')D_m(x-x')\right]\;,
\end{equation}
Here the first term  is the contribution of the zero mode, the second
term comes from the continuum of non-zero modes. We can
divide the continuum modes into symmetric and anti-symmetric subsets with
respect to the brane. Anti-symmetric modes vanish on the brane, so
they do not contribute to the Green's function at $z=0$. Since we are
calculating the gauge field on the brane, we can neglect
anti-symmetric modes in Eq. (\ref{summode}). The symmetric modes are
\begin{equation}
\label{solution}
A_m(z)=\e^{k\nu z} 
\sqrt{{m\over 2k}}{N_{\nu-1}({m\over k})J_{\nu}({m\over k}\e^{kz})
-J_{\nu-1}({m\over k})N_{\nu}({m\over k}\e^{kz})\over \sqrt{
N_{\nu-1}({m\over k})^2+J_{\nu-1}({m\over k})^2}}
\end{equation}
where
\[
\nu={n\over 2}+1\;.
\]
These continuum modes are normalized to $\delta(m-m')$ with the weight
(\ref{3+}). The retarded four-dimensional massive propagators entering
Eq.~(\ref{summode}) are conveniently written in coordinate
representation,
\begin{equation}
\label{10*}
D_m(x)=-{1\over 2\pi}\theta(t)\delta(\lambda^2)+{m\over
4\pi\lambda}\theta(t-|{\bf x}|)J_1(m\lambda),\;
\end{equation}
\[
\lambda=\sqrt{t^2-|{\bf x}|^2}\;.
\]
As a consistency check of our calculation, we point out that had one
boldly made use of the zero mode approximation, i.e., neglected the
continuum contribution in Eq.~(\ref{summode}), one would obtain from
Eq.~(\ref{4*}) that the induced gauge potential $A^0$ is a static
Coulomb potential, $A^0\propto 1/r$. This is not surprising, as
electrodynamics is effectively four-dimensional in the zero mode
approximation. 

The zero mode approximation is not justified when charges move outside
the brane, so we proceed to evaluate the complete Green's function. At
large distances, $t,|{\bf x}|\gg k^{-1}$, and well away from the light
cone, $\lambda\gg k^{-1}$, only relatively small masses $m$ are
relevant, and we obtain
\begin{equation}
\label{KK}
G_R(x,0,z')=-{\theta(t-|{\bf x}|)\over (2k)^{\nu-1}4\pi\lambda
\Gamma(\nu-1)}\e^{k\nu z'}\int_0^{\infty}dmm^{\nu}J_1(m\lambda)
J_{\nu}(m\xi')
\end{equation}
where we set $z=0$, as we are interested in the induced field on the
brane. Note that the first term in Eq.~(\ref{10*}) cancels out due to
completeness of the set of the modes (including the zero mode). 

The induced electromagnetic potential is obtained by plugging the
expression for the Green's function, Eq.~(\ref{KK}), into
Eq.~(\ref{4*}). The evaluation of the resulting integrals is somewhat
cumbersome, and we outline this step in Appendix. Let us present the
results for $A^0({\bf x},t)\equiv A^0(r,t)$ for $n=1$ and $n=2$.

$n=1:$
\begin{equation}
\label{9.}
A^0 = {ke\over 4\pi^2} \left[ {\sqrt{t^2-r^2}\over t^2} +
{1\over r} \left\{ {\rm arctg} \left({t+r\over t-r} \right)^{1/2}
- {\rm arctg} \left({t-r\over t+r} \right)^{1/2} \right\}\right]\;.
\end{equation}

$n=2:$
\begin{equation}
\label{9..}
A^0={ek\over 4\pi}\cdot {3t^2-r^2\over 2t^3}
\end{equation}
These expressions are valid inside the light cone $r<t$, whereas outside the
light cone $A^0$ is still equal to the Coulomb potential generated by
the charge that has been at rest on the brane at negative times,
\[
A^0={nek\over 8\pi}\cdot {1\over r}\;,
\]
the factor $nk/2$ coming from the normalization of the zero photon
mode. At both $n=1$ and $n=2$, the gauge potential $A^0$ and electric field
$E=-\d_r A^0$ are continuous on the light cone. Deep inside the light
cone, i.e., at $r\ll t$, the electric field gradually disappears,
\begin{equation}
\label{13*}
E\propto {r\over t^3}\;.
\end{equation}
The case of arbitrary even $n$ is also possible to treat analytically;
by making use of Eq.~(\ref{s5*}) we have checked that the same properties
(continuity of $A^0$ and $E$ across the light cone and the behavior
(\ref{13*}) at $r\ll t$) hold for all even $n$. The electromagnetic
field induced on the brane by escaping charges switches off in a
causal way.

We conclude that the electric charge non-conservation in the brane
world via the leakage of charged particles into extra dimensions is
consistent with effectively four-dimensional electrodynamics governing
the long distance interactions of charges residing on the brane. In
the setup we introduced in this paper, escaping charge induces
spherical electromagnetic wave on the brane; beyond this wave
electromagnetic field gradually disappears. This wave is a collective
effect of the photon zero mode and continuum modes.

The existence of continuum bulk modes with arbitrarily small
four-di\-men\-sional ``masses'' is crucial for this phenomenon, as it
makes the problem intrinsically five-dimensional even at large
distances, and in this way the 4-d Gauss' law obstruction to the
electric charge non-conservation is avoided. We expect that the
non-conservation of electric charge on the brane is possible in all
models with localized photon and continuum of electromagnetic modes in
the bulk that starts from zero four-dimensional ``mass''. These
properties are inherent, in particular, in the six-dimensional string
setup of Gherghetta and Shaposhnikov \cite{str3}; the localization
of a photon in the latter setup has been discussed recently by Oda
\cite{Oda}. On the other hand, those mechanisms of the photon
localization which do not incorporate the bulk continuum starting from
zero will not allow for the electric charge non-conservation in the
brane world; an example is the Dvali--Shifman setup
\cite{DvaliShifman}. It remains to be understood which possibility is
preferable from the point of view of string/M-theory.

May colored particles --- quarks --- escape from our brane too?
Certainly not in theories with confinement of color both on the brane
and in the bulk. However, if color is not confined in the bulk, one
may speculate on the following possibility. Consider a well separated
pair of quark and anti-quark on the brane, with color flux tube
stretching between them (fig. 1a). Suppose that the quark leaves the
brane (fig. 1b); if color is not confined in the bulk, this does not
cost very large energy even if the quark travels far away from the
brane. After the quark has moved away, there remains an object on the
brane (shown by dashed line in fig. 1c) which, from the
four-dimensional point of view, behaves as a color triplet state with
zero baryon number, electric and weak charges, etc. This object may
combine with anti-quark to form a free anti-quark colorless state
(again as viewed from four dimensions) with quantum numbers of
anti-quark (fig. 1d). It is worth trying to undertsand at a more
quantitative level whether these exotic objects may indeed exist in
models with infinite extra dimensions.

We are indebted to F.~Bezrukov, D.~Gorbunov, S.~Dimopoulos, G.~Dvali,
M.~Li\-ba\-nov, M.~Shaposhnikov, S.~Sibiryakov and S.~Troitsky for
helpful discussions.  This work is supported in part by RFBR grant
99-02-18410, by the CRDF grant 6603 and by Swiss Science Foundation,
grant 7SUPJ062239. The work of S.D. is also supported by the Russian
Academy of Sciences, JRP grant 37 and by ISSEP fellowship. The work of
P.T. is supported in part by the Swiss Science Foundation, grant
21-58947.99.

\newpage
\begin{center}
\epsfig{file=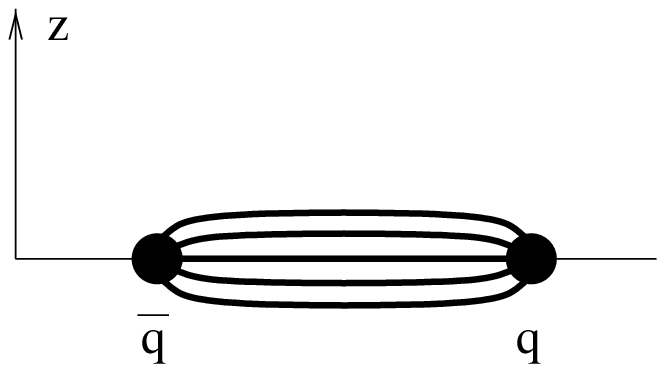}\\
Fig. 1a\\
\epsfig{file=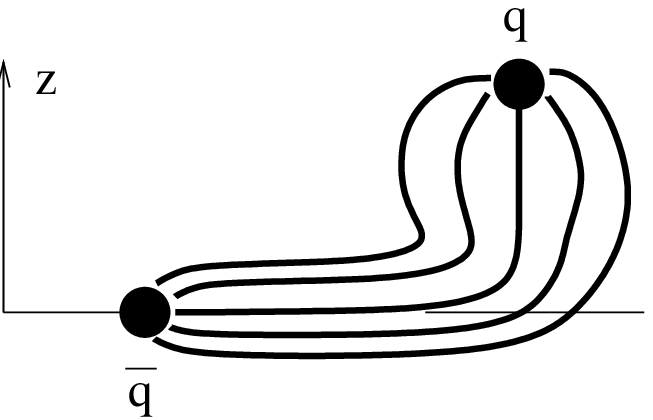}\\
Fig. 1b\\
\epsfig{file=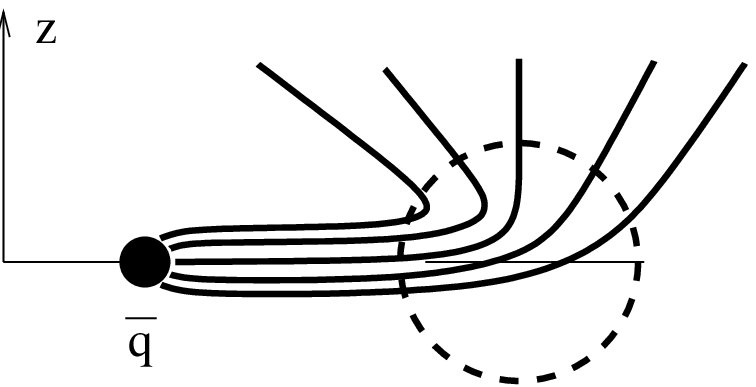}\\
Fig. 1c\\
\epsfig{file=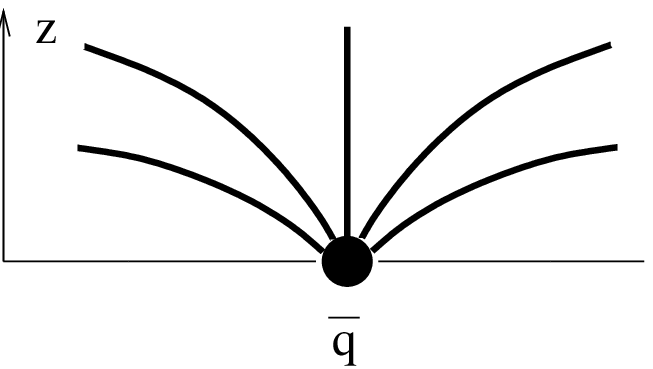}\\
Fig. 1d
\end{center}

\section*{ Appendix.}
The calculations of the integral entering Eq.~(\ref{KK}) are different
for even $n$ (integer $\nu$) and odd $n$ (half-integer $\nu$). Let us
begin with even $n$. One makes use of the relation
\[
m^{\nu}J_{\nu}(m\xi)=(-1)^{\nu}\xi^{\nu}\l{d\over\xi\d\xi}\r^{\nu}J_0(m\xi)
\]
and obtains
\begin{eqnarray}
\int_0^{\infty}dmm^{\nu}J_1(m\lambda)
J_{\nu}(m\xi)=(-1)^{\nu}\xi^{\nu}\l{d\over\xi\d\xi}\r^{\nu}
\int_0^{\infty}dmJ_1(m\lambda)
J_{0}(m\xi)\nonumber\\=(-1)^{\nu}\xi^{\nu}\l{d\over\xi\d\xi}\r^{\nu} 
{\theta(\lambda-\xi)\over\lambda}\;.
\end{eqnarray}
Hence, the Green's function is concentrated on the five-dimensional
light cone $\xi=\lambda$. Plugging this expression into
Eq. (\ref{KK}) and then Eq.~(\ref{4*}) one finds that the induced
field on the brane is given by the following integral 
\begin{eqnarray}
\label{integA}
A^0=-\alpha_e\int dt'\int_{1/k}^\infty {d\xi}(k\xi)^n
{\theta(t-t'-r)\over k^{2\nu-3}\lambda(t')^2}
{\delta(t'-\sqrt{\xi^2-k^{-2}})\over\sqrt{1-\l k\xi\r^{-2}}}\nonumber
\\
\times\xi^2\l
{d\over \xi d\xi}\r^{\nu}
\theta\l\lambda(t')-\xi\r\ \ \ \ \ \ 
\end{eqnarray}
where
\[
\alpha_e=\l-{1\over 2}\r^{\nu-1}{e\over 4\pi\Gamma\l \nu-1\r}
\]
and
\[
\lambda(t')^2=\l t-t'\r^2-r^2\;.
\]
We further simplify this integral by making use of the identity
\begin{equation}
\label{7*}
{d\over \xi d\xi}f(\xi)=\left.2{d\over 
d\beta}f(\sqrt{\xi^2+\beta})\right|_{\beta=0}
\end{equation}
which holds for arbitrary function $f(\xi)$. Then Eq. (\ref{integA})
takes the form
\begin{equation}
\begin{array}{cl}
\displaystyle A^0=\alpha_e k\l2{d\over 
d\beta}\r^{\nu-1}\int dt'\int_{0}^{\infty} d\xi&\displaystyle \xi^{n+2}
{\theta(t-t'-r)\over \lambda(t')^2}\nonumber \\
&\displaystyle \times{\delta(t'-\xi)\over \sqrt{\xi^2+\beta}}
\delta\l\lambda(t')-\sqrt{\xi^2+\beta}\r\biggr|_{\beta=0}
\end{array}
\end{equation}
where subleading in $\lambda/k$ terms have been omitted.
Now it is straightforward to perform the integration with the result
\begin{equation}
\label{s5*}
A^0(r,t)=\alpha_e k\left.\l2{d\over 
d\beta}\r^{\nu-1}{\xi_*^{n+2}\over t(\xi_*^2+\beta)}\right|_{\beta=0}
\end{equation}
where
\begin{equation}
\label{8*}
\xi_*={t^2-r^2-\beta\over 2t}\;.
\end{equation}
At $n=2$ Eq.~(\ref{s5*}) reduces to Eq.~(\ref{9..}).

Let us now evaluate the integral entering
Eq.~(\ref{summode})
at odd $n$
(half-integer $\nu$). We make use of the relation
\[
m^{l+1/2}J_{l+1/2}(m\xi)=(-1)^l\sqrt{2\over\pi}\xi^{l+1/2} 
\l{d\over\xi\d\xi}\r^{l}
{\sin{m\xi}\over\xi}
\]
Then 
\begin{eqnarray}
\int_0^{\infty}dmm^{\nu}J_1(m\lambda)
J_{\nu}(m\xi')=(-1)^{l}\sqrt{2\over\pi}\xi^{\nu}\l{d\over\xi\d\xi}\r^{l}
\int_0^{\infty}dmJ_1(m\lambda){\sin{m\xi}\over\xi}
\nonumber\\=(-1)^{l}\sqrt{2\over\pi}\xi^{\nu}\l{d\over\xi\d\xi}\r^{l} 
{\theta(\lambda-\xi)\over\lambda\sqrt{\lambda^2-\xi^2}}
\end{eqnarray}
This expression is again plugged into Eq.~(\ref{KK}) and then
Eq.~(\ref{4*}). The integration over $t'$ is performed by making use
of the trick (\ref{7*}), and we find
\begin{equation}
\label{last}
A^0=\alpha_o k\left.\l2{d\over 
d\beta}\r^{\nu-1/2}\int_0^{\xi_*}{d\xi\xi^{n+2}\over\lambda(\xi)^2
\sqrt{\lambda(\xi)^2-\xi^2-\beta}}\right|_{\beta=0}
\end{equation}
where
\[
\alpha_o=\l-{1\over2}\r^{\nu-3/2}{e\over 4\pi^{3/2}\Gamma(\nu-1)}\;.
\]
At $n=1$ the integral (\ref{last}) can be evaluated explicitly,
and one obtains Eq.~(\ref{9.}).

\end{document}